\documentclass{ws-procs9x6}
\usepackage{epsfig}
\usepackage{subfigure}

\begin{document}

\title{Deeply Virtual Compton Scattering at Hera II (H1 results)}

\author{B. Roland}

\address{I.I.H.E., Universit\'e libre de Bruxelles \\
Boulevard du Triomphe, CP 230, \\ 
1050 Bruxelles, Belgium\\ 
E-mail: broland@ulb.ac.be}

\maketitle

\abstracts{New results on the Deeply Virtual Compton Scattering process
$\gamma^* p \to \gamma p$ (DVCS) from H1 experiment at the $e p$ collider HERA
are presented, using data collected during the year 2004 and corresponding to
an integrated luminosity of 39.7 pb$^{-1}$. The DVCS cross section is measured 
as a function of the photon virtuality, $Q^2$, the $\gamma^* p $ c.m.s. energy $W$ 
and differentially in the momentum transfer squared at the proton vertex, $t$,
in the kinematic range $6.5 < Q^2 < 80 $ GeV$^2$, $30 < W < 140$ GeV and $|t| < 1$ GeV$^2$.
The results are found to be in good agreement with the published H1 and ZEUS measurements 
and compared to NLO QCD calculations based on Generalized Parton Distributions (GPD) 
and to predictions from Color Dipole Model.}

\section{Introduction}

The Deeply Virtual Compton Scattering process  $\gamma^* p \to \gamma p$ (DVCS) 
is the hard diffractive scattering of a virtual photon off a proton.
It has a clear experimental signature identical to that of Bethe-Heitler (BH) process.
This latter being purely electromagnetic and involving only elastic proton form factors
is precisely known and can therefore be bin by bin subtracted, 
the integration over the azimuthal angle ensuring that the interference term between the 2 processes vanishes.
Furthermore, BH can be used to study the detector response in the analysis kinematic range.   
DVCS is of particular interest to test the description perturbative Quantum Chromodynamics (pQCD) gives
of the colorless exchange, as it does not suffer, as the diffractive vector meson (VM) production,
of the uncertainties on the VM wave function. In the hard regime, for $Q^2 \gg M_p^2$ and $|t| \ll Q^2$,
the DVCS amplitude factorizes \cite{Collins:1998be,Radyushkin:2000ap} into the convolution of coefficient functions, pQCD calculable,
and universal functions describing the structure of the proton in the non forward kinematic domain,
the Generalized Parton Distributions or GPD. These functions extend the usual parton distribution functions (pdf) 
by taking into account the difference between the longitudinal momentum fractions of the emitted and absorbed partons,
the skewedness $\xi$, and the momentum transfer squared at the proton vertex, $t$.


\section{Data Analysis}

These measurements of the DVCS cross section are based on the data collected by the H1 detector during the year 2004, 
with HERA running with positrons colliding protons of energy 27.5 and 920 GeV respectively, 
and correspond to an integrated luminosity of 39.7 pb$^{-1}$. 
To enhance the ratio of DVCS events w.r.t. BH ones, the photon is required to be detected
in the forward\footnote{The forward region corresponds to the outgoing proton direction} or central region of the H1 detector,
with a transverse momentum $P_T > 2$ GeV, while the scattered positron is detected in the backward region, with an energy
$E > 15$ GeV. To ensure the elastic selection and reduce the proton dissociation background, the absence of activity
in the forward detectors is required.   
To extract the $e p \to e p \gamma$ cross section, the BH and inelastic DVCS backgrounds are subtracted bin by bin
and the data are corrected for trigger efficiency, detector acceptance and initial state photon radiation 
using the Monte Carlo simulation Milou \cite{Perez:2004ig}. The $\gamma^* p \to \gamma p$ cross section is then extracted 
using the equivalent photon approximation.  

\section{Results}

\begin{figure} [ht]
\vspace*{0.7cm}
\epsfxsize=5.5cm
\begin{picture}(60,80)
\put(90,-5){\epsfbox{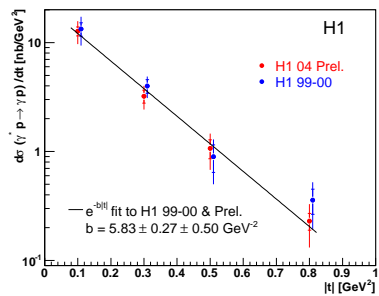}}
\end{picture}
\caption{DVCS $\gamma^* p \to \gamma p$ differential cross section in $t$ at $Q^2$ =  8 GeV$^2$.\label{Meas_t}}
\vspace*{-0.2cm}
\end{figure}

Figure \ref{Meas_t} shows the $\gamma^* p \to \gamma p$ differential cross section in $t$ at $Q^2$ =  8 GeV$^2$.

The H1 04 preliminary measurements are found to be in agreement with the H1 published ones \cite{Aktas:2005ty}, 
while fitting the $t$ dependence of the combined data  with a function of the form $e^{-b|t|}$ 
reduces the statistical uncertainty on the $t$ slope, leading to $b = 5.83 \pm 0.27 \pm 0.50$ GeV$^{-2}$.

\begin{figure} [ht]
\vspace*{0.6cm}
\epsfxsize=5.5cm
\begin{picture}(60,80)
\subfigure{
\put(10,-5){\epsfbox{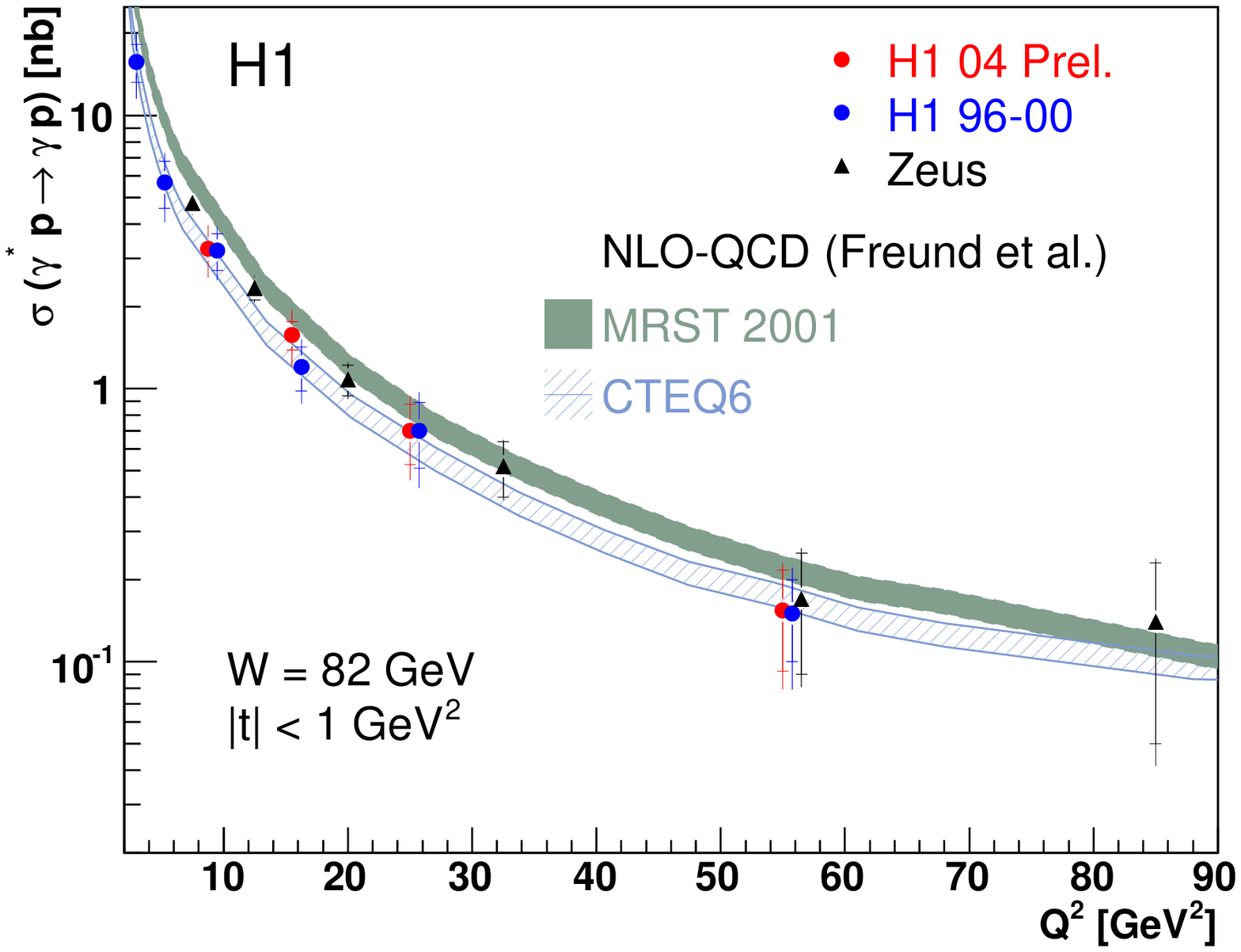}}
\label{NLOQCD_Q2}}
\subfigure{
\put(160,-5){\epsfbox{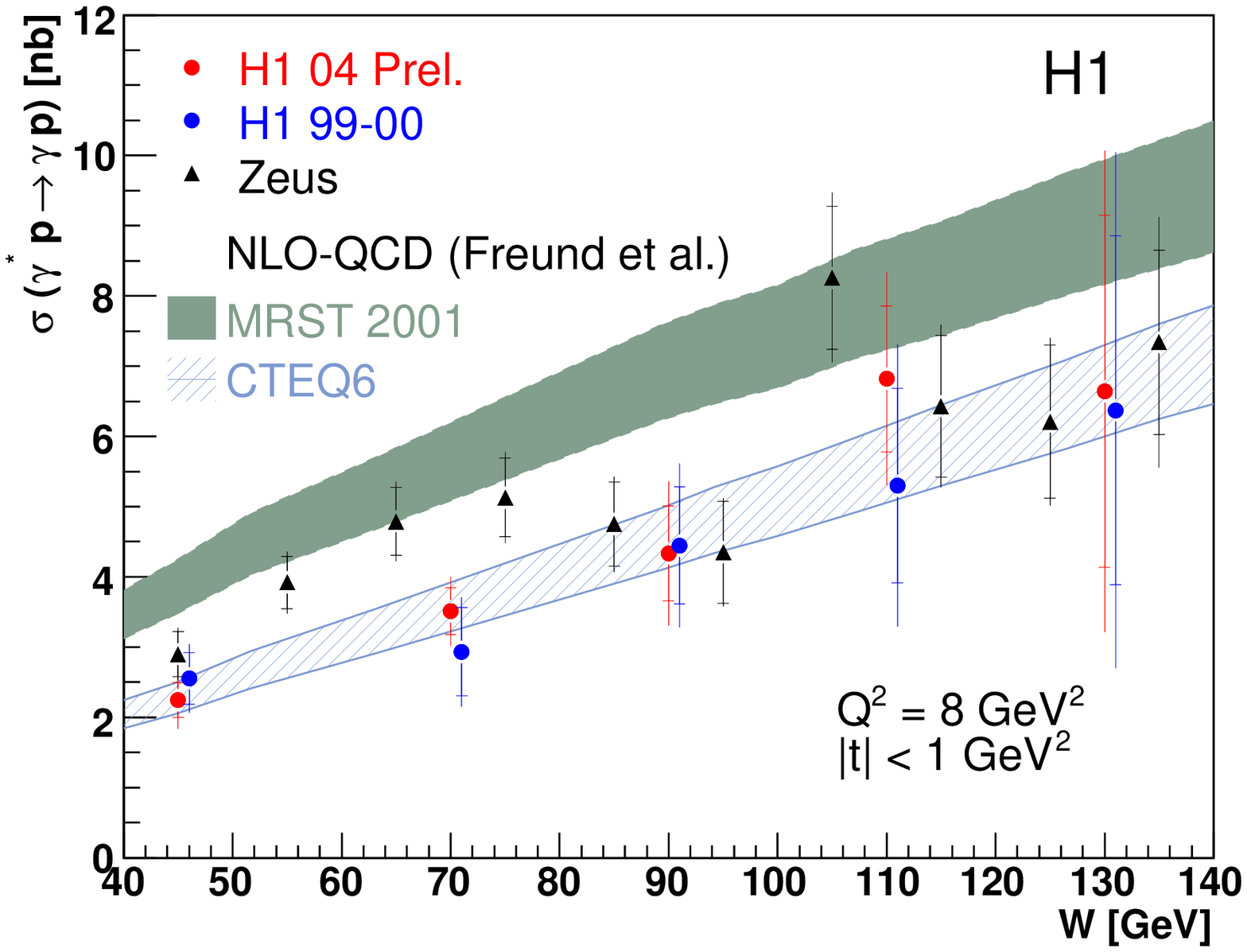}}
\label{NLOQCD_W}}
\put(60,95){(a)}
\put(225,95){(b)}
\end{picture}
\caption{DVCS $\gamma^* p \to \gamma p$ cross section as a function of $Q^2$ for $W$ = 82 GeV (a) and of $W$ for $Q^2$ =  8 GeV$^2$ (b). 
The H1 04 preliminary measurements are shown together with the published H1 and ZEUS results 
and compared to NLO QCD predictions using MRST 2001 and CTEQ6 pdf. The error band associated to the different curves 
corresponds to the uncertainty on the $t$ slope measurement.\label{NLOQCD}}
\end{figure}

Figures \ref{NLOQCD_Q2} and \ref{NLOQCD_W} show the $\gamma^* p \to \gamma p$ cross section as a function of $Q^2$ and $W$,
for $W$ = 82 GeV and $Q^2$ =  8 GeV$^2$ respectively. Fitting the $Q^2$ dependence of the combined H1 results 
with a function of the form $(1/Q^2)^n$ gives $n = 1.52 \pm 0.07 \pm 0.04$, while fitting the $W$ dependence with a function of the form
$W^{\delta}$ gives $\delta = 1.00 \pm 0.16 \pm 0.22$. This value, comparable to that of diffractive $J/\Psi$ VM production, indicates 
the presence of the hard regime. The H1 04 preliminary measurements are found to be in agreement with the H1 \cite{Aktas:2005ty} 
and ZEUS \cite{Chekanov:2003ya} published results, while the H1 combined fit reduces 
the statistical uncertainty on n and $\delta$ parameters.
Figures \ref{NLOQCD_Q2} and \ref{NLOQCD_W} also compare the data points to the NLO QCD predictions 
by Freund and McDermott \cite{Freund:2002qf}. For the DGLAP region $|x| > \xi$, the usual pdf $q(x;\mu^2)$ from MRST2001 and CTEQ6 
are used to parametrize the GPD $\mathcal{H}$ whose contribution is dominant in the probed low x region, 
the $t$ dependence being factorized in an exponential behaviour. 
At a given scale $\mu^2$, the quark singlet and gluons distributions are respectively given by 
$\mathcal{H}^q(x,\xi,t;\mu^2) = q(x;\mu^2)e^{-b|t|}$, $\mathcal{H}^g(x,\xi,t;\mu^2) = x g(x;\mu^2)e^{-b|t|}$. 
In the ERBL region, for $|x| < \xi$, the quark singlet and gluons distributions are parametrized by simple analytic
functions, satisfying GPD symmetry properties and continuation to the DGLAP region. The $Q^2$ and $\xi$ dependences are 
then generated dynamically through the evolution equations. The QCD predictions are in good agreement with the data, in both
shape and normalisation, whose uncertainty is reduced by the measurement of the $t$ slope parameter $b$ 
and inferior to that affecting the pdf over all the $Q^2$ and $W$ range of the measurement. The comparison between data
and theoretical estimations shows no need for intrinsic skewedness.

\begin{figure} [ht]
\vspace*{0.8cm}
\epsfxsize=5.5cm
\begin{picture}(60,80)
\subfigure{
\put(10,-5){\epsfbox{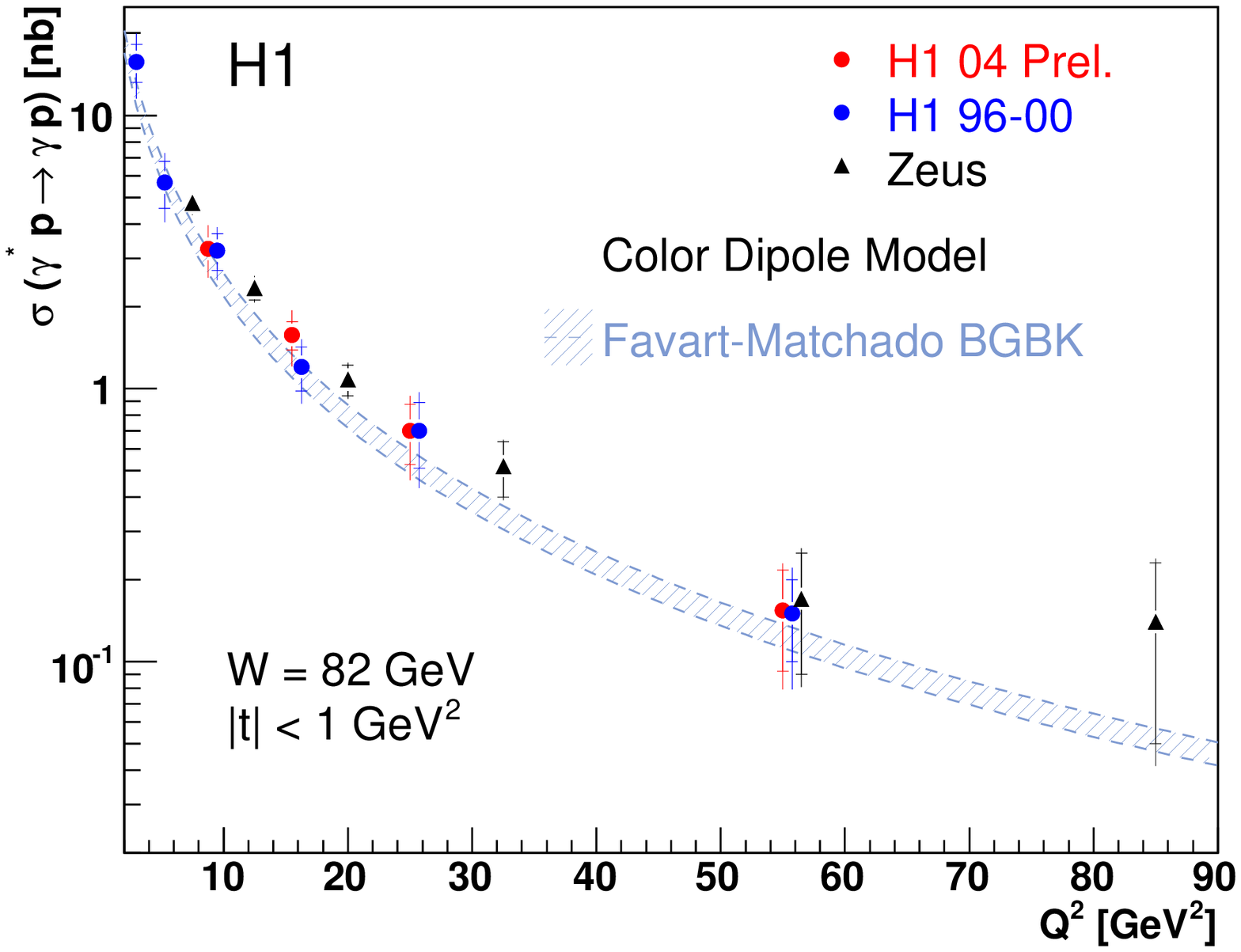}}
\label{Dipole_Q2}}
\subfigure{
\put(160,-5){\epsfbox{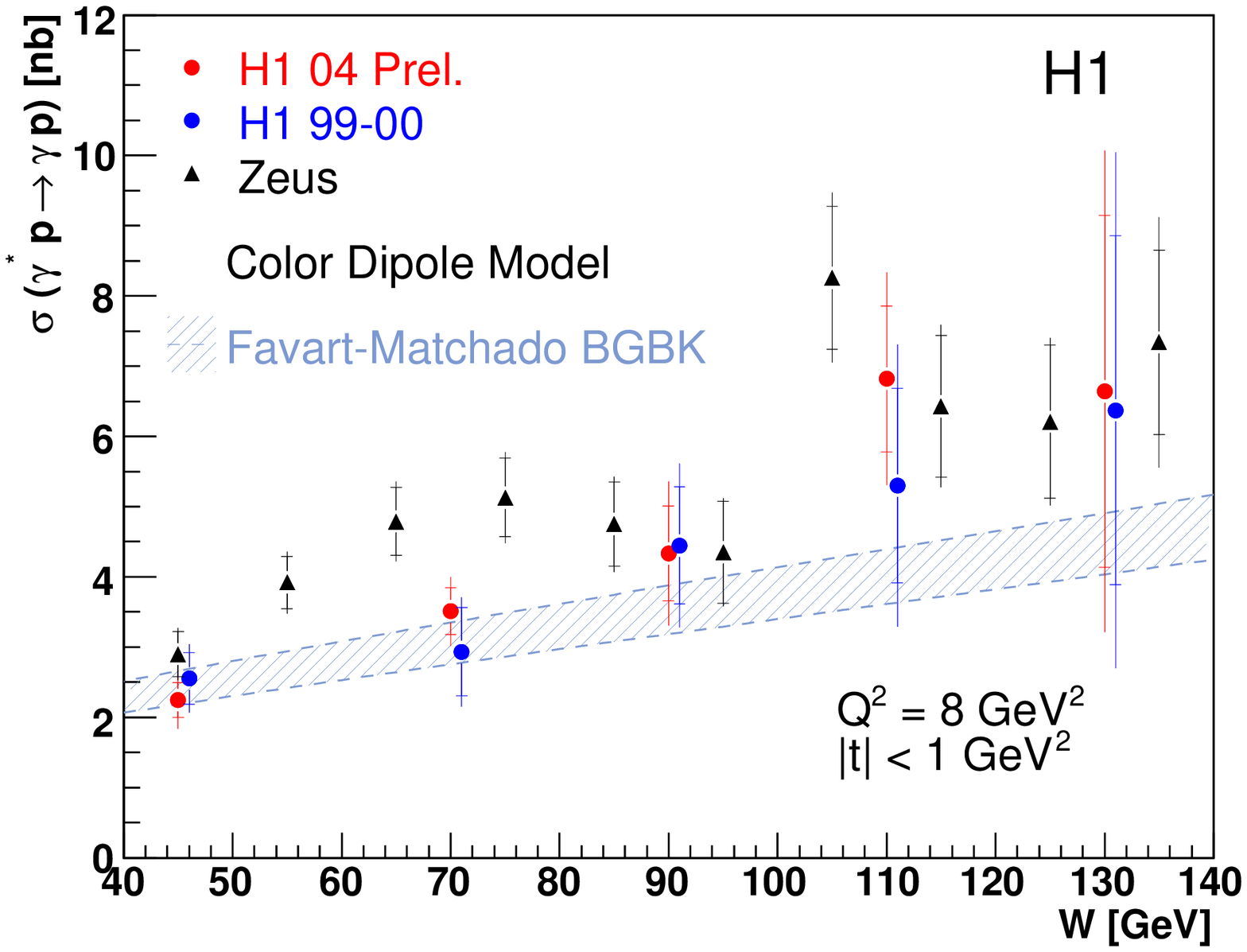}}
\label{Dipole_W}}
\put(60,95){(a)}
\put(225,95){(b)}
\end{picture}
\vspace*{0.2cm}
\caption{DVCS $\gamma^* p \to \gamma p$ cross section as a function of $Q^2$ for $W$ = 82 GeV (a) and of $W$ for $Q^2$ =  8 GeV$^2$ (b). 
The H1 04 preliminary measurements are shown together with the published H1 and ZEUS results 
and compared to predictions from Color Dipole Model. The error band associated to the different curves 
corresponds to the uncertainty on the $t$ slope measurement.\label{Dipole}}
\end{figure}

Figures \ref{Dipole_Q2} and \ref{Dipole_W} compare the data points to Color Dipole Model prediction
by Favart and Machado \cite{Favart:2004uv}, which is found to give a reasonable description of the measurement 
by applying to DVCS a saturation approach including a DGLAP evolution of the dipole. 

\section{Summary}

The H1 04 preliminary results are found to be in agreement with the H1 published ones, NLO QCD predictions and Color Dipole Model
expectations. The H1 combined fit has allowed to reduce the statistical uncertainty on the parameters $b, n$ and $\delta$.

\end{document}